\documentclass[english, aps, pre, reprint, notitlepage, superscriptaddress, longbibliography]{revtex4-1}
\usepackage[T1]{fontenc}
\usepackage[utf8]{inputenc}
\usepackage{amssymb}
\usepackage{amsmath}
\usepackage{amstext}
\usepackage{graphicx}
\usepackage{esint}
\usepackage{babel}
\usepackage{xcolor}
\usepackage[normalem]{ulem}

\usepackage[breaklinks=true]{hyperref}
\usepackage{breakcites}
\usepackage{breakurl}

\begin{document}

\title{The distinguishable-particle lattice model of glasses in three dimensions}

\author{Bo Li}
\affiliation{School of Science, Harbin Institute of Technology (Shenzhen), Shenzhen, 518055, China}

\author{Chun-Shing Lee}
\affiliation{Department of Applied Physics, Hong Kong Polytechnic University, Hung Hom, Hong Kong, China}
\affiliation{School of Science, Harbin Institute of Technology (Shenzhen), Shenzhen, 518055, China}

\author{Xin-Yuan Gao}
\affiliation{Department of Applied Physics, Hong Kong Polytechnic University, Hung Hom, Hong Kong, China}
\affiliation{Department of Physics, The Chinese University of Hong Kong, Shatin, New Territories, Hong Kong, China}

\author{Hai-Yao Deng}
\email{dengh4@cardiff.ac.uk}
\affiliation{School of Physics and Astronomy, Cardiff University, 5 The Parade, Cardiff CF24 3AA, Wales, UK}

\author{Chi-Hang Lam}
\email{c.h.lam@polyu.edu.hk}
\affiliation{Department of Applied Physics, Hong Kong Polytechnic University, Hung Hom, Hong Kong, China}

\begin{abstract}
The nature of glassy states in realistic finite dimensions is still under fierce debates. Lattice models can offer valuable insights and facilitate deeper theoretical understanding. Recently, a disordered-interacting lattice model with distinguishable particles in two dimensions (2D) has been shown to produce a wide range of dynamical properties of structural glasses, including the slow and heterogeneous characteristics of the glassy dynamics, various fragility behaviors of glasses, and so on. These findings support the usefulness of this model for modeling structural glasses.
An important question is whether such properties still hold in the more realistic three dimensions.
In this study, we aim to extend the distinguishable-particle lattice model (DPLM) to three dimensions (3D) and explore the corresponding glassy dynamics. 
Through extensive kinetic Monte Carlo simulations, we found that the 3D DPLM exhibits many typical glassy behaviors, such as plateaus in the mean square displacement of particles and the self-intermediate scattering function, dynamic heterogeneity, variability of glass fragilities, and so on, validating the effectiveness of DPLM in a broader realistic setting.
The observed glassy behaviors of 3D DPLM appear similar to its 2D counterpart, in accordance with recent findings in molecular models of glasses.
We further investigate the role of void-induced motions in dynamical relaxations and discuss its relation to dynamic facilitation.
As lattice models tend to keep the minimal but important modeling elements, they are typically much more amenable to analysis. Therefore, we envisage that the DPLM will benefit future theoretical developments, such as the configuration tree theory, towards a more comprehensive understanding of structural glasses. 
\end{abstract}

\maketitle

\section{Introduction}
After a rapid cooling from the liquid state, many materials can form glasses, with a drastic increase of viscosity while maintaining their structural disorder~\cite{donthbook, debenedetti2001, berthier2011review}. In order to unveil the nature of the glassy state and the glass transition, it is important to understand the hallmarks and the origin of the slow dynamics in glasses. 
Experimental and molecular dynamics (MD) simulations are valuable to examine the microscopic dynamics of particles in a glass~\cite{kob1995, kob1999}. However, it appears challenging to obtain a thorough picture from microscopic data, mostly due to the complexity and the lack of structural order in glassy systems.
Various theoretical models to obtain valuable insights and guidance for further experimental explorations are under intensive studies.

Thermodynamic theories for describing the slow glassy dynamics are typically coarse-grained, capturing the trapping of the systems in local minima of the energy landscape, where the energy barriers hinder the relaxation of particles at low temperatures~\cite{berthier2011review}. Dynamic facilitation theories based on the kinetically constrained models incorporate particle rearrangements, which naturally explain many properties of glassy dynamics, such as dynamical heterogeneity~\cite{garrahan2011review}.
Recent advances in MD simulations coupled with the swap Monte Carlo method can reach experimental glass transition temperatures; the results point to the important role of dynamic facilitation, which are more prominent at low temperatures~\cite{scalliet2022}.

To bridge the gap between abstract theories and realistic observations, it would be valuable to develop effective lattice models which are amenable to analysis. A notable example in statistical physics is the Ising model for studying magnetism and the corresponding phase transitions~\cite{kersonhuang1987}.
The kinetically constrained models belong to this category, most of which have a facilitation rule by model construction. For example, in the Fredrickson and Andersen (FA) model, the particle in a local region is allowed to evolve only when it is facilitated by the presence of mobile regions around it~\cite{fredrickson1984}. Other lattice models for glasses which mimic hard-sphere systems are also proposed~\cite{biroli2001, ciamarra2003, krzakala2008, nishikawa2020}; they typically possess geometric frustration that can give rise to glassy behaviors. In this work, we focus on a lattice model called the distinguishable-particle lattice model (DPLM)~\cite{zhang2017}. In contrast to the abovementioned models, the DPLM is featured by random interactions that is quenched in the configuration space, which can naturally model glassy systems with particles having distinct properties or frustration states~\cite{zhang2017}. One salient feature of the DPLM is that it exhibits an emergent facilitation behavior~\cite{zhang2017}.
The DPLM is less coarse-grained in comparison to typical kinetically constrained models, as a particle represents an atom or a tightly bounded group of atoms while, in contrast, a spin state in the FA model represents the density of a local region. Therefore, the DPLM offers more flexibility to investigate different glassy phenomena.
Like the DPLM with emergent facilitation, a new lattice model also with such a property has been proposed recently~\cite{hasyim2023}.

The two-dimensional (2D) DPLM has been shown to produce a wide range of dynamic characteristics of glasses, such as plateaus in the mean square displacement of particles and the self-intermediate scattering function, violation of the Stokes-Einstein relation, dynamic heterogeneity and so on.
Notably, it exhibits emergent dynamic facilitation behaviors~\cite{zhang2017}.
More recently, the 2D DPLM have successfully reproduced even more intricate thermodynamic and kinetic phenomena observed in realistic glassy materials, such as the kinetic fragility and the Kovacs' paradox~\cite{lulli2020, lee2020}.
As an interacting lattice gas model, it also allows close examinations of the effects of void-induced motions, which has been directly observed in colloidal systems and numerical simulations of hard disks~\cite{yip2020}.

Albeit successful in explaining many dynamical phenomena of glasses, the DPLM was primarily studied in two-dimensional lattices~\cite{zhang2017, lulli2020, lee2020, lee2021, gao2022, gopinath2022}, although a related model was simulated in three-dimensions (3D) as well~\cite{gao2023Kauzmann}. As is known, the spatial dimension plays a vital role in many statistical mechanical systems~\cite{justin2002}. Drastic differences of glassy dynamics between 2D and 3D have been reported~\cite{flenner2015}. But follow-up studies argued that the differences can be accounted for by the Mermin-Wagner long-wavelength fluctuations, except which there is no fundamental difference~\cite{shiba2016, weeks2017}. 
On the other hand, previous studies have identified some peculiar mechanical and thermodynamic behaviors of glasses in 2D~\cite{Flenner2019, berthier2019zero}. Therefore, there are still a large room to explore the role of dimensionality in glasses.

We should also emphasize that generalizing lattice models defined in a certain dimension to other dimensions could be highly non-trivial~\cite{newman1999, jack2016, ritort2003review, garrahan2011review}.
Hence, it is essential to study and provide clear evidence that the a lattice model for glasses can be extended to three dimensions while maintaining key glass-like properties.
Therefore, it is important to extend the DPLM to the unexplored 3D case, which is the main focus of this work. 
Through extensive kinetic Monte Carlo simulations, we found that the 3D DPLM is able to capture many interesting glassy phenomena as in the 2D case, which further validates the effectiveness of DPLM. The results demonstrate a qualitative similarity between the 3D and 2D models.
Whether there exists other intricate differences of glassy behaviors among different dimensions remains to be explored in the further. Our work lays a solid foundation for such purposes.
It will also be interesting to study its dynamics based on theoretical approaches, such as mode-coupling theory~\cite{Gotzebook} and the configuration tree theory~\cite{lam2018tree,deng2019}. One immediate benefit of the obtained quantitative results here is to fit some key parameters in the configuration tree theory.
Lastly, we remark that the vibrational modes of glasses are not considered in the DPLM, due to its lattice nature. As a result, some dynamical features, such as short-time vibrational behaviors, are not directly captured by the DPLM.

The remaining part of the article is organized as follows. In Sec.~\ref{sec:model}, we introduce the DPLM model and the methods for conducting simulations. We then discuss the results in Sec.~\ref{sec:results} and conclude the paper in Sec.~\ref{sec:conclusion}.

\section{Model}\label{sec:model}
The DPLM was first introduced in~\cite{zhang2017}, while a simplified set of particle interactions is proposed in~\cite{lee2020} which keeps all the interesting glassy properties. We will therefore adopt the DPLM model of~\cite{lee2020} in what follows. It is defined by $N$ distinguishable particles living on lattices. We will focus on 3D simple cubic lattices of sizes $L^3$, and consider periodic boundary conditions. Each particle $i$ has a label $s_i \in \{1,2,...,N\}$ indicating its identity. Each lattice site can be occupied by at most one particle; denote the occupation number on site $i$ as $n_i \in \{0, 1\}$. Site $i$ is said to be occupied by a void if $n_i = 0$.

Consider two nearest neighboring sites $i$ and $j$ on the lattice. If and only if both sites are occupied by particles (with the corresponding labels $s_i$ and $s_j$), there will be an interaction energy $V_{s_i, s_j}$ incurred. One crucial feature of the DPLM is that the interaction energy at the bond between these two sites $i$ and $j$ depends on the identities $s_i$ and $s_j$ of the particles sitting on these sites.   
The total energy of the system is
\begin{equation}
E(\boldsymbol{s}) = \sum_{\langle i,j \rangle } V_{s_i, s_j} n_i n_j,
\end{equation}
where the summation is over all bonds of the lattice. 

We further assume that the interaction energy $V_{kl}$ is a quenched random variable drawn independently and identically from a certain distribution $g(V)$. Note that $k, l \in \{ 1, 2, ..., N \}$ are particle labels, but not site indices. In contrast to the usual spin glass systems where the interaction energy $J_{ij}$ of bond $\langle i,j \rangle$ are quenched random variables, the interaction energy on the corresponding bond in DPLM can change over time, as the particle labels $s_i$ and $s_j$ on sites $i$ and $j$ are dynamical variables. On the other hand, whenever sitting on adjacent sites, two particles $k$ and $l$ will always incur the same energy $V_{kl}$. Following Ref.~\cite{lee2020}, we consider that the interaction energies $V_{kl}$ are drawn from the mixture of a uniform distribution on the interval $[V_0, V_1]$ and a Dirac delta, with the probability density
\begin{equation}\label{eq:gV_def}
g(V) = \frac{G_0}{V_1 - V_0} + (1 - G_0) \delta(V - V_1), \quad V_0 \leq V \leq V_1,
\end{equation}
where $G_0 \in [0, 1]$ is a thermodynamic parameter which has an influence on the fragility of the glass system~\cite{lee2020}. 

Here, we elaborate more on the interaction energy distribution.
In this work, we consider $V_0 = -0.5, V_1 = 0.5$, which set the energy scales of the attractive bonds and repulsive bonds. 
The interaction energy $V_{kl}$ between particles $k$ and $l$ has a probability $G_0$ to be drawn from a uniform distribution between $V_0$ and $V_1$; the uniform distribution is chosen only for modeling simplicity. On the other hand, $V_{kl}$ has a probability $1 - G_0$ to take that value $V_1$, which corresponds to a bond excited state. Such a mixture distribution can be motivated by the two-state model~\cite{angell2000}. 
The above uniform-plus-delta distribution of $g(V)$ allows us to model glasses of various fragilities due to the interplay between energetics and entropy of the bonds at low temperatures~\cite{lee2020}; also see Sec.~\ref{sec:fragility} for more details.
Other choice of $g(V)$ is also possible, depending on the properties of the materials. For instance, a mixture distribution of two Dirac delta with supports at $V_0$ and $V_1$ has been considered in Ref.~\cite{ong2022}, which exhibits a fragile-to-strong transition.
It would also be interesting to design suitable interaction energy distribution to study glasses with repulsion and attraction at different distances~\cite{dsawson2000}, as well as exploring the fragilities of network-forming systems~\cite{SaikaVoivod2001, ciarella2019pnas}.
We remark that the randomness of the interaction energy is essential for the producing glassy behaviors in the DPLM; setting $G_0 = 0$ will creates a simple lattice gas with uniform interactions.

Interestingly, the equilibrium statistics of the DPLM are exactly solvable, based on which a direct initialization method to quickly equilibrate the system has been developed~\cite{zhang2017}. At a certain temperature $T$, each particle on the lattice can hop to an unoccupied neighboring site with the hopping rate
\begin{equation}\label{eq:w_def}
w = \begin{cases}
w_0 \exp [ -(E_0 + \Delta E) / (k_B T) ], 
\quad & \Delta E > 0, \\
w_0 \exp [ - E_0 / (k_B T) ], 
\quad & \Delta E \leq 0 
\end{cases}
\end{equation}
where $\Delta E$ is the change in system energy due to the hop, $k_B$ is the Boltzmann constant (set as $k_B = 1$), and $w_0$ and $E_0$ are two tunable parameters.
The prefactor $w_0$ can be considered as the attempt rate of the proposed transition, while $E_0$ is an offset to the hopping energy barrier.
Without loss of generality, we set $w_0 = 10^6$ and $E_0 = 0$ if not specified otherwise. The kinetic Monte Carlo dynamics defined in Eq.~(\ref{eq:w_def}) satisfies detailed balance. 
Note that when a particle at site $i$ (suppose its label is $s_i = A$) hops to an unoccupied site $j$, it carries its label to site $j$. That is, we have $s_i = A, n_i=1$ and $n_j = 0$ before hopping, while $n_i=0$ and $s_j = A, n_j = 1$ after hopping.

Equivalently, the dynamics of the systems can also be characterized by the movement of voids in the reverse direction of the particle hopping. Therefore, it is straightforward to observe and study void-induced particle motions and their collective behaviors in the DPLM, which have been shown to play an important role in glass dynamics in both colloidal systems and MD simulations~\cite{yip2020}. We also reiterate that the DPLM is a lattice model without considering the vibrational modes of glasses; therefore, the dynamical phenomena related to vibrations are typically not captured.

\section{Results}\label{sec:results}
Here, we simulate the dynamics of the DPLM on a 3D cubic lattice with $L=32$ at various temperatures according to the kinetic Monte Carlo update rule of Eq.~(\ref{eq:w_def}), starting from the corresponding equilibrium states. We will typically consider the regime of high particle density, i.e., $\phi := N/L^3  \lessapprox 1$. For later convenience, we define the void density as $\phi_v = 1 - \phi$. In the regime where both $\phi_v$ and $T$ are small so that the systems exhibit glassy dynamics, the rejection-free method proposed in Ref.~\cite{zhang2017} provides efficient numerical simulations. Unless specified otherwise, the results below are averaged over five independent runs with different disordered realizations of the couplings $\{ V_{kl} \}$.

\subsection{Mean Square Displacement and Diffusion Coefficient}
We firstly report the behaviors of particle mean square displacement (MSD) defined as $\text{MSD}(t) = \langle | \boldsymbol{r}_l(t) - \boldsymbol{r}(0) |^2 \rangle$ where $\boldsymbol{r}_l(t)$ is the position of particle $l$ at time $t$ and $\langle \cdots \rangle$ denotes the average over time and ensemble. 

\begin{figure}
\hspace*{-1em}\includegraphics[scale=0.9]{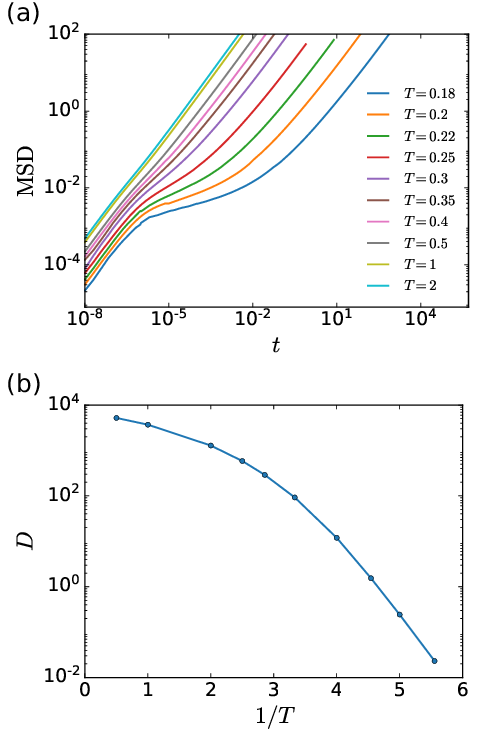}
\caption{(a) Particle mean square displacement of the DPLM on a 3D cubic lattice with $L=32$ as a function of time in log-log scale, calculated at various temperatures.
Void density $\phi_v = 0.01$ is considered. Other parameters for the simulations are $G_0 = 1, V_0 = -0.5, V_1 = 0.5, w_0 = 10^6, E_0 = 0$. The results are averaged over five independent runs with different realizations of the couplings. (b) Diffusion coefficient $D$ of the particles plotted as a function of the inverse temperature $1/T$. 
\label{fig:msd}}
\end{figure}

We consider a uniform distribution of the interaction energies by setting $G_0 = 1$ in Eq.~(\ref{eq:gV_def}). 
The results of MSD are shown in Fig.~\ref{fig:msd}(a). At low temperatures, there appears to be a plateau in the MSD curve, which is a signature of glassy behaviors. 

In the large time limit $t \to \infty$, all the slopes of the MSD curves at different temperatures approach unity, indicating a diffusive behavior with $\text{MSD} \propto t$. In the diffusive regime, we measure the diffusion coefficient defined as
\begin{equation}
D = \frac{1}{2d} \lim_{t \to \infty} \frac{ \text{MSD}(t) }{t},
\end{equation}
where $d = 3$ is the spatial dimension considered. In practice, we determine the diffusive regime by requiring a large enough $t$ such that $\text{MSD}(t) > 1$ and $\text{MSD}(t) \propto t^{\gamma}$ with $\gamma > 0.96$. The result is shown in Fig.~\ref{fig:msd}(b). Similar to the 2D cases, the diffusion coefficients in 3D DPLM also exhibit super-Arrhenius behaviors.

\subsection{Self-intermediate Scattering Function}
Another important dynamical observable is the self-intermediate scattering function (SISF), which can be directly compared to neutron or X-ray scattering experiments. It is defined as
\begin{equation}\label{eq:Fs_def}
F_s(\boldsymbol{q}, t) = \langle e^{\mathrm{i} \boldsymbol{q} \cdot ( \boldsymbol{r}_l(t) - \boldsymbol{r}_l(0) )} \rangle,
\end{equation}
where $\boldsymbol{q}$ is a wave vector related to the wave length $\lambda$ as $q = 2 \pi / \lambda$.
As we are interested in the length scale of particle hopping, we choose the wave vector $\boldsymbol{q}$ corresponding to a small wave length $\lambda = 2$. 

\begin{figure}
\hspace*{-1em}\includegraphics[scale=0.9]{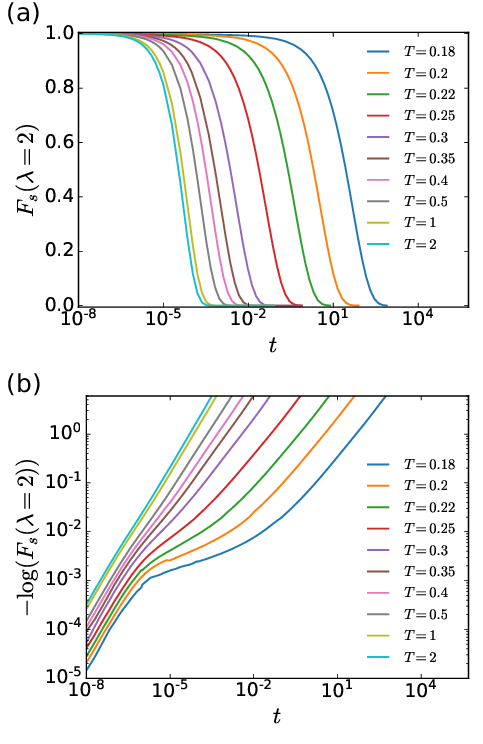}
\caption{Dynamical evolution of self-intermediate scattering function $F_s(q, t)$ of the DPLM on a 3D cubic lattice. The wave vector is chosen such that the wave length $\lambda = 2$. The system parameters are the same as those in Fig.~\ref{fig:msd}.
(a) $F_s(q, t)$ versus $t$ in the linear-log scale. (b) $-\log ( F_s(q, t) )$ versus $t$ in the log-log scale.
\label{fig:sisf}}
\end{figure}

As shown in Fig.~\ref{fig:sisf}(a), the SISF exhibits a two-step decay with a plateau before decaying to zero at low temperatures, which indicates a slow relaxation dynamic of the DPLM. As the modes of atomic vibration are not considered in our lattice model, it cannot be directly compared with relaxations of realistic glasses, but is analogous to the SISF of inherent structures~\cite{liao2001v2}.

At large time, the decay of $F_s(q, t)$ can be fitted by the Kohlrausch-Williams-Watts (KWW) stretched exponential function $A \exp(-(t / \tau_\alpha)^{\beta})$, where $\tau_\alpha$ is a relaxation time and $\beta$ is the stretching exponent. This can be made transparent when we plot $-\log ( F_s(q, t) )$ as a function of $t$ in the log-log scale, as shown in Fig.~\ref{fig:sisf}(b); the linear behaviors of the curves at large time indicate a good approximation of the KWW function with $A \approx 1$. 

\begin{figure}
\hspace*{-1em}\includegraphics[scale=0.9]{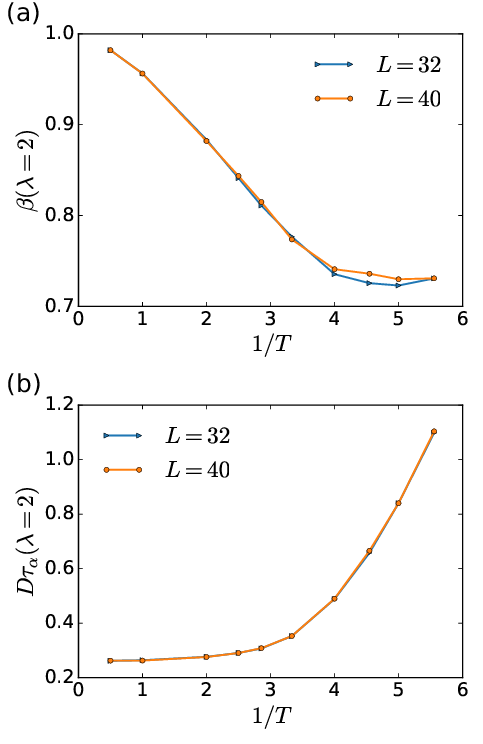}
\caption{(a) Stretching exponent $\beta$ as a function of the inverse temperature $1/T$. The system parameters are the same as those in Fig.~\ref{fig:msd}, except that two system sizes ($L=32$ and $L=40$) are considered. (b) The behavior of the product of diffusion coefficient and the $\alpha$-relaxation time $D \tau_\alpha$, where $D$ is obtained in Fig.~\ref{fig:msd}(b) and $\tau_\alpha$ is extracted from Fig.~\ref{fig:sisf}(a). 
\label{fig:beta_and_Dtau}}
\end{figure}

Various physical quantities can be extracted from the SISF. As usual, the $\alpha$-relaxation time $\tau_\alpha$ is defined as the time at which $F_s(q, t)$ drops to $1/e$. Observing the large-time characteristics of SISF in Fig.~\ref{fig:sisf}(b), we calculate the stretching exponent $\beta$, by finding the slope of $\log_{10} [ -\log ( F_s(q, t) ) ]$ against $\log_{10} (t)$ around $\tau_\alpha$. The behavior of $\beta$ at different temperatures is shown in Fig.~\ref{fig:beta_and_Dtau}(a), which is close to $1$ at high temperatures and decreases when the temperature is lowered, which is another indication of glassy dynamics at lower temperatures. The slight increase of $\beta$ when $T$ gets lower than $T = 0.2$ for the system with $L=32$ seems to be a finite size effect, which is alleviated when a larger system size is considered. 

We also show the behavior of $D \tau_\alpha$ in Fig.~\ref{fig:beta_and_Dtau}(b) when the temperature is varied; it exhibits a violation of the Stokes-Einstein relation $D \tau_\alpha$ = \text{constant}, as expected for glasses.
The Stokes-Einstein violation has been attributed to dynamic heterogeneity and the interplay between particle hopping and non-hopping motions~\cite{yamamoto1998, kawasaki2017, shiba2019, kawasaki2013}. The DPLM lacks non-hopping motions and this may explain why the magnitude of violation found here is smaller than typical ones from MD simulations.

\subsection{Four-point Correlation function and Dynamic Heterogeneity}
One intriguing feature of glass-forming systems is the emergence of dynamic heterogeneity at low temperatures, where some regions in the system relax much slower than others~\cite{berthier2011book}.
This feature makes glasses drastically different from liquids, even though both systems have very similar disordered structures. 

Here, we examine the spatial and temporal heterogeneity of particle dynamics in the 3D DPLM numerically, following the procedures in Ref.~\cite{zhang2017}. We first define an ``overlap'' function 
\begin{equation}
c_l(t, 0) = e^{ \mathrm{i} \boldsymbol{q} \cdot ( \boldsymbol{r}_l(t) - \boldsymbol{r}_l(0) )  },
\end{equation}
which probes the movement of particle $l$ from time $0$ to time $t$ at a length scale $2\pi / q$. 
A dynamical overlap field can be defined based on this overlap function as
\begin{equation}
c(\boldsymbol{r}; t, 0) = \sum_l c_l(t, 0) \delta( \boldsymbol{r} - \boldsymbol{r}_l(0) ),
\end{equation}
which is a measure of overlap (of length scale $2 \pi / q$) in a region near $\boldsymbol{r}$.

The spatial fluctuations of the particle overlap field are naturally captured by its correlation function
\begin{equation}
G_4 ( \boldsymbol{r}, t ) = \langle c(\boldsymbol{r}; t, 0) c(\boldsymbol{0}; t, 0) \rangle - \langle  c(\boldsymbol{0}; t, 0) \rangle^2.
\end{equation}
We then define the corresponding susceptibility as $\chi_4(t) = \int \mathrm{d} \boldsymbol{r} G_4 ( \boldsymbol{r}, t )$, which can be interpreted as the size of the correlated clusters in the relaxation dynamics~\cite{berthier2011reviewv2}. 

\begin{figure}
\hspace*{-1em}\includegraphics[scale=0.5]{ 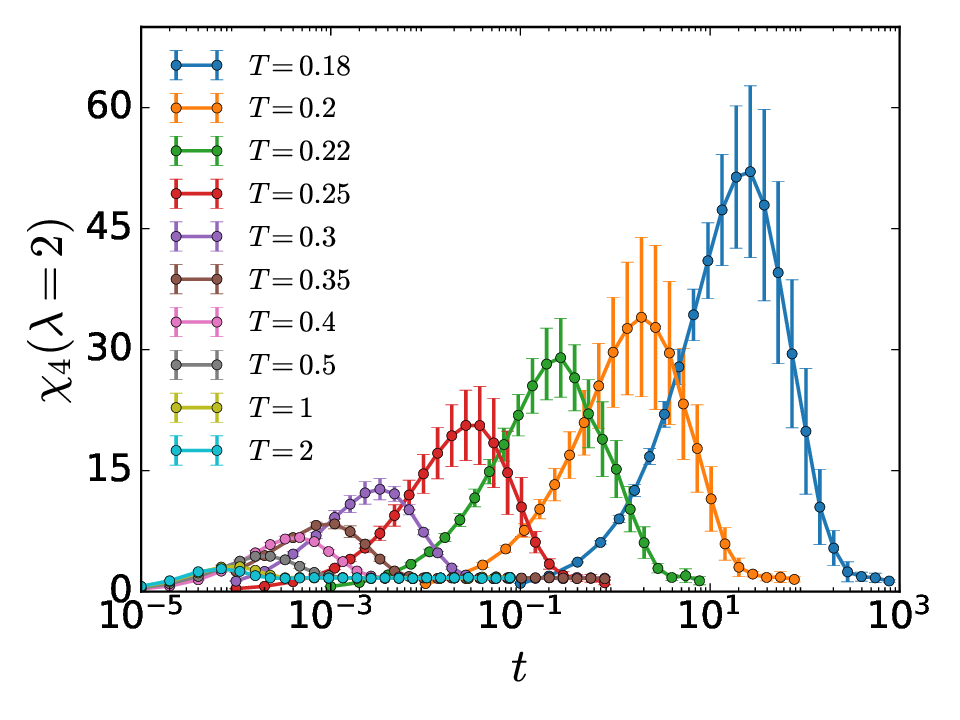}
\caption{Evolution of the dynamic susceptibility $\chi_4$ at different temperatures. The error bar indicates one standard deviation across the five disordered realizations. The system parameters are the same as those in Fig.~\ref{fig:msd}. 
\label{fig:chi4}}
\end{figure}

The behaviors of the dynamic susceptibility $\chi_4(t)$ are shown in Fig.~\ref{fig:chi4}, which exhibit a peak for each temperature. When the temperature decreases, the peaked time of $\chi_4(t)$ increases rapidly, whose scale is comparable to that of structural relaxation; the peak height of $\chi_4(t)$ also increases but only mildly. These behaviors indicate that an increasing correlation of the dynamical clusters is developed when the temperature gets lower and the dynamics slow down, supporting the existence of dynamic heterogeneity in such systems. These phenomena are also commonly observed in various glassy systems~\cite{berthier2011book} as well as in the 2D DPLM.
The increasing peak height of $\chi_4(t)$ with decreasing temperatures suggests an increasing correlation length scale. It will be interesting to further examine the growth of this length scale and possible scaling behaviors with decreasing temperatures and increasing systems sizes in future studies. 

\subsection{Emergent Facilitation Behaviors}
Unlike the kinetically constrained models, the DPLM is not constructed with an explicit facilitation rule in the dynamics. In spite of this, it exhibits emergent dynamic facilitation behaviors, which become transparent from the motions of voids.

\begin{figure}
\hspace*{-1em}\includegraphics[scale=0.8]{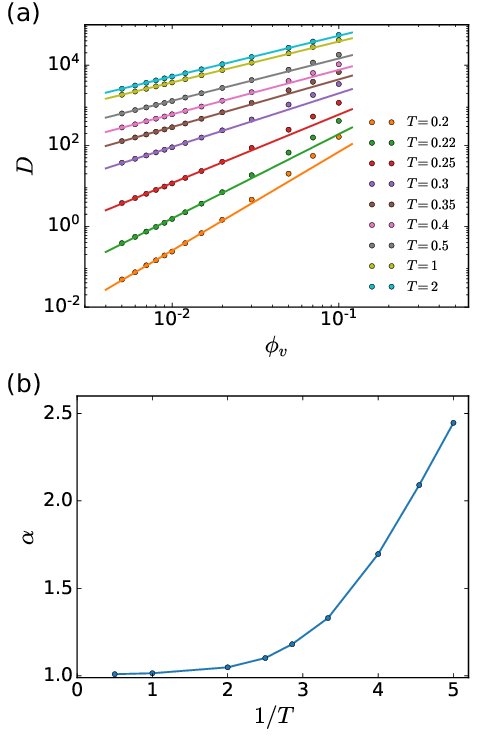}
\caption{(a) Diffusion coefficient $D$ of the 3D DPLM, as a function of the void density $\phi_v$ (in the log-log scale) at various temperatures. Other system parameters are $G_0 = 1, V_0 = -0.5, V_1 = 0.5, w_0 = 10^6, E_0 = 0$. A linear fit to the data $\{ ( \log_{10} \phi_v, \log_{10} D ) \}$ for small void densities with $\phi_v \leq 0.05$ is performed to extract the exponent $\alpha$. (b) Scaling exponent $\alpha$ versus $1/T$.
\label{fig:vary_phiv}}
\end{figure}

Following Ref.~\cite{zhang2017}, we examine how void density $\phi_v$ impacts the system dynamics by contrasting the diffusion coefficient $D$ against $\phi_v$. The results are shown in Fig.~\ref{fig:vary_phiv}(a). It exhibits a linear relation in the log-log scale, especially in the regime of low $\phi_v$, which suggests a diffusion coefficient power law relation
\begin{equation}
D \sim \phi_v^{\alpha}. \label{eq:D_power_law}
\end{equation}
We therefore perform a linear fit to the data $\{ ( \log_{10} \phi_v, \log_{10} D ) \mid \phi_v \leq 0.05 \}$ for various temperatures, where the slope corresponds to the exponent $\alpha$. The result is shown in Fig.~\ref{fig:vary_phiv}(b). 
At high temperatures, it is observed that $\alpha \simeq 1$, indicating individual movements of voids~\cite{palmer1990}. The scaling exponent $\alpha$ increases as the temperature decreases, indicating that the movements of voids become more and more collective (under longer and longer observation time). For example, for $\alpha \simeq 2$, pairs of coupled voids dominate the system movements; as the chance that two voids meet is of the order $\phi_v^2$, we expect that $D \sim \phi_v^2$.

\begin{center}
\begin{figure}
\includegraphics[scale=0.5]{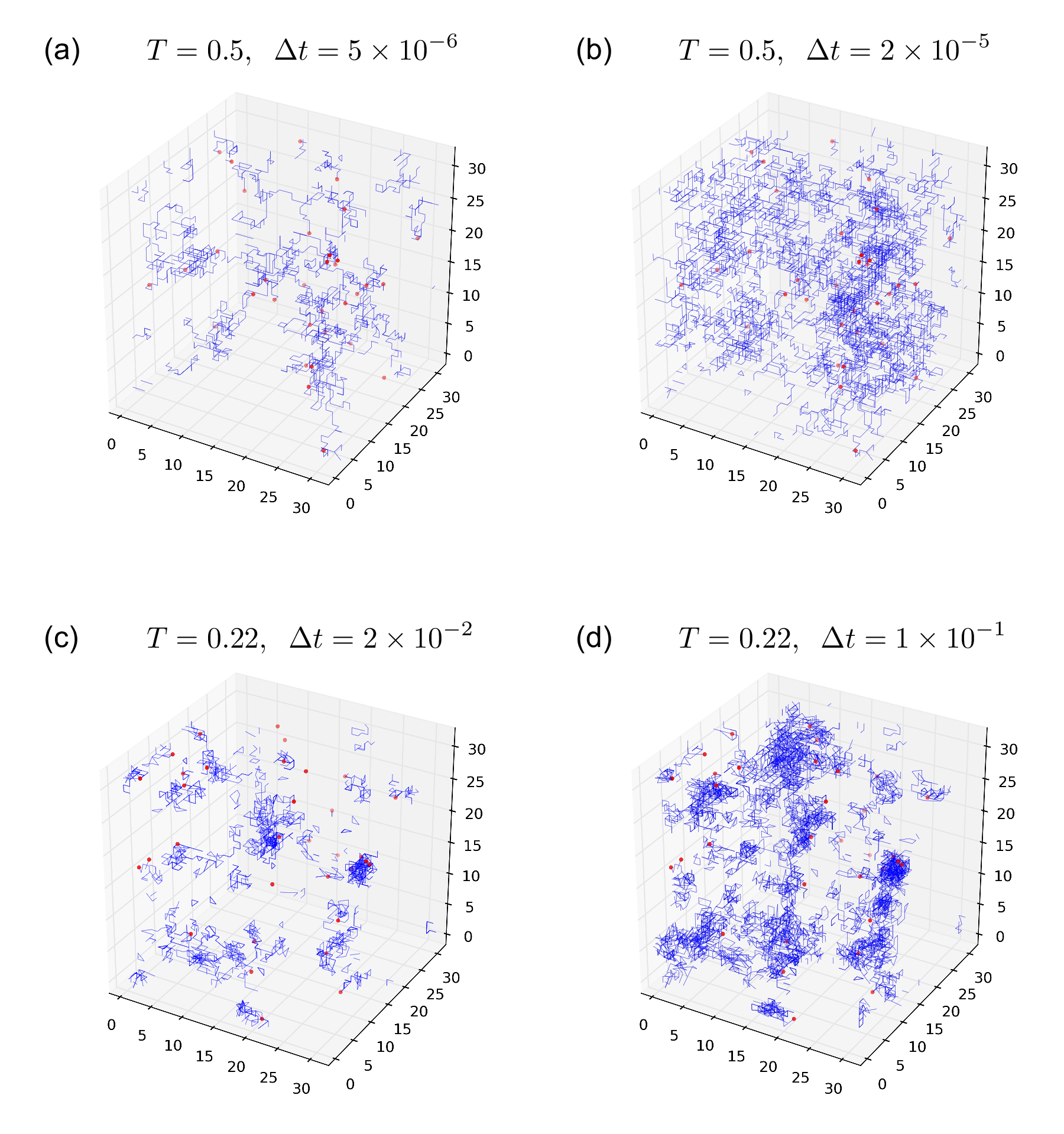}
\caption{Movements of particles in a 3D DPLM with $L = 32$ and $\phi_v$ = 0.001 within a time window $\Delta t$ at temperature $T$. Other system parameters are the same as those in Fig.~\ref{fig:msd}.
The blue lines indicate the particle trajectories, where particle positions are recorded at the time interval of $\Delta t / 10$. The red circles indicate the initial positions of the void, which are initiators of the particle movements.
Panels (a) and (b) correspond to a system at temperature $T=0.5$, while panels (c) and (d) correspond to a system at temperature $T=0.22$.
In both (a) and (c), the SISF $F_s(\lambda=2)$ drops to about 0.97 after time $\Delta t$.
}
\label{fig:trajs}
\end{figure}
\end{center}

To make this picture more explicit, we visualize the particle movements of a 3D DPLM system in Fig.~\ref{fig:trajs}. At a high temperature $T = 0.5$ corresponding to an exponent $\alpha \simeq 1$, it can be seen in Fig.~\ref{fig:trajs}(a) that individual voids can induce sustained particle movements. Conversely, at a low temperature $T = 0.22$ corresponding to an exponent $\alpha \simeq 2$, it can be observed in Fig.~\ref{fig:trajs}(c) that the regions exhibiting more particle hopping activities (or equivalently, more void hopping activities) often consist of groups of nearby voids, while isolated voids tend to remain trapped in the vicinity of their initial positions.
This can be considered as a facilitation behavior, where a void becomes more mobile in the vicinity of another void. As such mobile groups of voids are heterogeneously distributed in space and time, the emergent facilitation behavior directly contributes to dynamic heterogeneity in glasses. Fig.~\ref{fig:trajs}(d) showcases the heterogeneous particle hopping activities at a low temperature $T=0.22$ for a longer time window, which is in sharp contrast to those at a high temperature $T=0.5$ as shown in Fig.~\ref{fig:trajs}(b).

\onecolumngrid
\twocolumngrid

\clearpage

\subsection{Fragility of glasses}\label{sec:fragility}
The kinetic fragility of glasses describes how rapidly the dynamics slow down when temperature decreases~\cite{bohmer1993, angell1995, wang2006}, which can be indicated by the behaviors of viscosity, particle diffusion coefficient, and so on.
Fragile glasses possess a more drastic dynamic slowdown, while the opposite holds for strong glasses.
Remarkably, it has been shown in the previous study of 2D DPLM~\cite{lee2020} that glasses of different fragilities can be obtained by varying the parameter $G_0$ defining the distribution of the random interactions in Eq.~(\ref{eq:gV_def}), as well as the parameter $E_0$ determining a hopping barrier offset in Eq.~(\ref{eq:w_def}). 
In particular, a small $G_0$ results in more fragile glasses~\cite{lee2020}. 
Physically, for the DPLM with a small $G_0$, most particle bonds have a high interaction energy with $V = V_1$, while a small number of bonds attain a lower energy with $V < V_1$.
As temperature decreases, the system has an increasing probability to be trapped in states with more low-energy pairings. Such states are not easy to access as they have a small entropy, but once attained (at a low enough temperature), they are exceptionally stable, resulting in a dramatic slowdown in the dynamics and corresponding to a high fragility~\cite{lee2020}.

\begin{figure}
\centering
\includegraphics[scale=0.55]{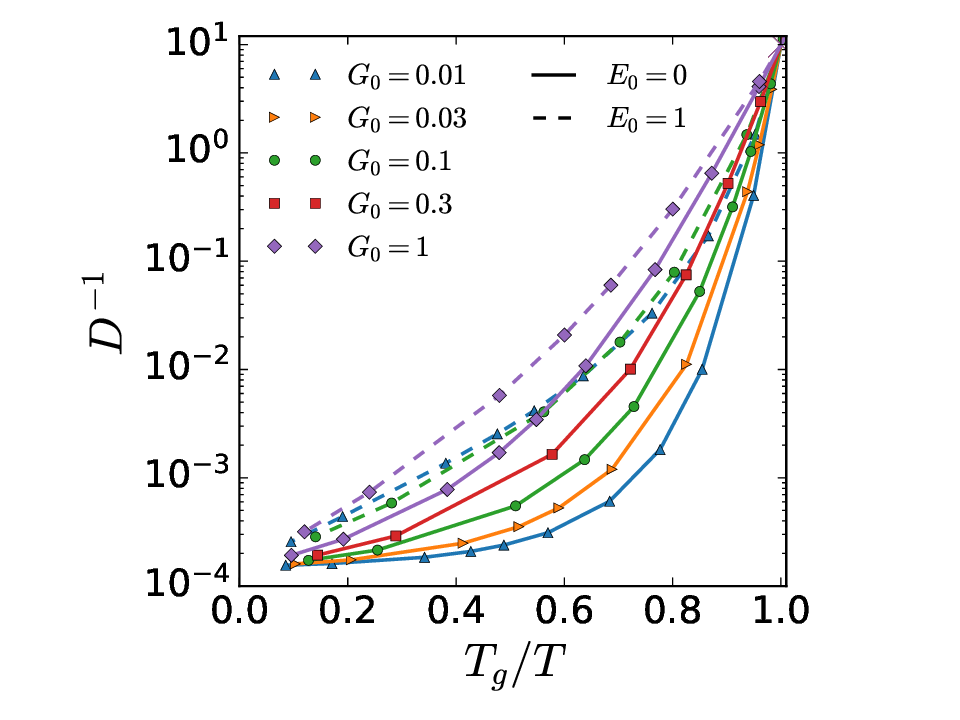}
\caption{Kinetic Angell plot of $D^{-1}$ against $T_g / T$ for different $G_0$ (represented by different colors) and different $E_0$ (represented by different line styles). Here, $T_g$ is defined as the temperature at which the particle diffusion coefficient $D$ becomes $D_r = 10^{-1}$.}
\label{fig:fragility}
\end{figure}

Here, we examine whether a similar phenomenon exists in 3D DPLM, by plotting $D^{-1}$ (as a measure of dynamic slowdown) against $T / T_g$, where $T_g$ is a glass ``transition'' temperature defined at which the particle diffusion coefficient $D$ becomes $D_r = 10^{-1}$. The result is shown in Fig.~\ref{fig:fragility}, which indicates that the glasses become more fragile as $G_0$ decreases, in accordance with the observation in 2D DPLM.
We also observe in Fig.~\ref{fig:fragility} that increasing the kinetic parameter $E_0$ in Eq.~(\ref{eq:w_def}) creates glassy systems that are less fragile, similar to the 2D case.

\section{Conclusion}\label{sec:conclusion}
In this work, we extended the DPLM to three dimensions and explored the glassy dynamics of such systems through extensive kinetic Monte Carlo simulations.
At low temperatures, both the mean square displacement of particles and the self-intermediate scattering function exhibit a plateau, which are signatures of glassy behaviors.

Dynamic heterogeneity is observed by considering the dynamic susceptibility of a certain particle overlap field, which reveals an increasing length scale of the dynamic correlation when the temperature decreases.
The increasing dynamic heterogeneity at lower and lower temperatures can be straightforwardly accommodated in the picture of dynamic facilitation observed in the DPLM. Individual voids become increasingly trapped when the temperature drops; the system dynamics are more and more dominated by pairs of voids, triplets of voids, etc, which are heterogeneously distributed in space and time. Interestingly, dynamic facilitation is an emergent property in the DPLM, which differs from more coarse-grained models such as the kinetically constrained models. 

We also observed various glass fragilities of the DPLM by tuning the parameter $G_0$ which defines the distribution of the interaction energy. 
All the glassy phenomena possessed by the 3D DPLM that have been examined in this work display no profound differences from the 2D counterpart.
The qualitative similarity of the elementary dynamical behaviors of glasses between the 3D DPLM and the 2D case is in accordance with the recent numerical studies on molecular models of glasses~\cite{shiba2016, weeks2017}.
In future works, it will be interesting to investigate other potential intricate differences between 2D and 3D glasses and their physical origin through the lens of DPLM.
Equilibrium statistics of the DPLM for both 2D and 3D are exactly available~\cite{zhang2017}. This has enabled us to directly generate 3D DPLM equilibrium states in the present work. In the future, we will also study its dynamics based on theoretical approaches, such as the configuration tree theory~\cite{lam2018tree, deng2019}.

The DPLM assumes void-induced particle hopping motions. Voids of sizes comparable to the particles are in general rare in glasses. However, a generalized form of void called quasivoid each consisting of neighboring free-volume fragments of a combined size comparable to a particle has been recently observed in experiments on colloidal glasses~\cite{yip2020}. This supports the void-induced dynamics assumed in this work.

It appears challenging to directly verify the diffusion coefficient power law in Eq.~(\ref{eq:D_power_law}) with respect to the void density in MD simulations because one cannot alter the void or quasivoid density without significantly impacting other system properties. Nevertheless, Eq.~(\ref{eq:D_power_law}) as reported in~\cite{zhang2017} has motivated an analogous diffusion coefficient power law in recent MD simulations of a molecular partial-swap system in which only a selected set of particles called swap-initiator can swap with each other or with the regular particles in the neighborhood~\cite{gopinath2022}. Noting that a void-induced hop of a particle can equivalently be considered as a swap between a void and a particle, Eq.~(\ref{eq:D_power_law}) then suggests a power-law with respect to the density of swap initiators, which is well verified in both MD simulations and modified DPLM simulations in which partial swaps dominate the dynamics.  In particular, the power law exhibited in the MD simulations is based on a system of polydisperse repulsive spheres which is more realistic than lattice models. The successful prediction of the power law in MD simulations thus supports the possible physical relevance of the DPLM in realistic glasses.

\section*{Author contributions}
C.H.L. and H.Y.D. conceived the project.
X.Y.G. wrote the computer programs.
B.L. conducted the numerical simulations, performed the analyses and wrote the initial version of the manuscript with support from C.S.L..
All authors discussed the methods and results and contributed to the final manuscript.

\section*{Conflicts of interest}
There are no conflicts to declare.

\begin{acknowledgments}
We acknowledge support from the start-up funding from Harbin Institute of Technology, Shenzhen (Grant No. 20210134), the Shenzhen Start-Up Research Funds (Grant No. BL20230925) the National Natural Science Foundation of China (Grant No. 12205066  and 11974297), and Hong Kong GRF (Grant No. 15303220).
\end{acknowledgments}

\bibliography{glass}

\begin{thebibliography}{50}%
\makeatletter
\providecommand \@ifxundefined [1]{%
 \@ifx{#1\undefined}
}%
\providecommand \@ifnum [1]{%
 \ifnum #1\expandafter \@firstoftwo
 \else \expandafter \@secondoftwo
 \fi
}%
\providecommand \@ifx [1]{%
 \ifx #1\expandafter \@firstoftwo
 \else \expandafter \@secondoftwo
 \fi
}%
\providecommand \natexlab [1]{#1}%
\providecommand \enquote  [1]{``#1''}%
\providecommand \bibnamefont  [1]{#1}%
\providecommand \bibfnamefont [1]{#1}%
\providecommand \citenamefont [1]{#1}%
\providecommand \href@noop [0]{\@secondoftwo}%
\providecommand \href [0]{\begingroup \@sanitize@url \@href}%
\providecommand \@href[1]{\@@startlink{#1}\@@href}%
\providecommand \@@href[1]{\endgroup#1\@@endlink}%
\providecommand \@sanitize@url [0]{\catcode `\\12\catcode `\$12\catcode
  `\&12\catcode `\#12\catcode `\^12\catcode `\_12\catcode `\%12\relax}%
\providecommand \@@startlink[1]{}%
\providecommand \@@endlink[0]{}%
\providecommand \url  [0]{\begingroup\@sanitize@url \@url }%
\providecommand \@url [1]{\endgroup\@href {#1}{\urlprefix }}%
\providecommand \urlprefix  [0]{URL }%
\providecommand \Eprint [0]{\href }%
\providecommand \doibase [0]{http://dx.doi.org/}%
\providecommand \selectlanguage [0]{\@gobble}%
\providecommand \bibinfo  [0]{\@secondoftwo}%
\providecommand \bibfield  [0]{\@secondoftwo}%
\providecommand \translation [1]{[#1]}%
\providecommand \BibitemOpen [0]{}%
\providecommand \bibitemStop [0]{}%
\providecommand \bibitemNoStop [0]{.\EOS\space}%
\providecommand \EOS [0]{\spacefactor3000\relax}%
\providecommand \BibitemShut  [1]{\csname bibitem#1\endcsname}%
\let\auto@bib@innerbib\@empty
\bibitem [{\citenamefont {Donth}(2001)}]{donthbook}%
  \BibitemOpen
  \bibfield  {author} {\bibinfo {author} {\bibfnamefont {E}~\bibnamefont
  {Donth}},\ }\href@noop {} {\emph {\bibinfo {title} {The glass transition:
  relaxation dynamics in liquids and disordered materials}}},\ Vol.~\bibinfo
  {volume} {48}\ (\bibinfo  {publisher} {Springer},\ \bibinfo {year}
  {2001})\BibitemShut {NoStop}%
\bibitem [{\citenamefont {Debenedetti}\ and\ \citenamefont
  {Stillinger}(2001)}]{debenedetti2001}%
  \BibitemOpen
  \bibfield  {author} {\bibinfo {author} {\bibfnamefont {Pablo~G.}\
  \bibnamefont {Debenedetti}}\ and\ \bibinfo {author} {\bibfnamefont
  {Frank~H.}\ \bibnamefont {Stillinger}},\ }\bibfield  {title} {\enquote
  {\bibinfo {title} {Supercooled liquids and the glass transition},}\ }\href
  {\doibase 10.1038/35065704} {\bibfield  {journal} {\bibinfo  {journal}
  {Nature}\ }\textbf {\bibinfo {volume} {410}},\ \bibinfo {pages} {259--267}
  (\bibinfo {year} {2001})}\BibitemShut {NoStop}%
\bibitem [{\citenamefont {Berthier}\ and\ \citenamefont
  {Biroli}(2011)}]{berthier2011review}%
  \BibitemOpen
  \bibfield  {author} {\bibinfo {author} {\bibfnamefont {Ludovic}\ \bibnamefont
  {Berthier}}\ and\ \bibinfo {author} {\bibfnamefont {Giulio}\ \bibnamefont
  {Biroli}},\ }\bibfield  {title} {\enquote {\bibinfo {title} {Theoretical
  perspective on the glass transition and amorphous materials},}\ }\href@noop
  {} {\bibfield  {journal} {\bibinfo  {journal} {Rev. Mod. Phys.}\ }\textbf
  {\bibinfo {volume} {83}},\ \bibinfo {pages} {587} (\bibinfo {year}
  {2011})}\BibitemShut {NoStop}%
\bibitem [{\citenamefont {Kob}\ and\ \citenamefont {Andersen}(1995)}]{kob1995}%
  \BibitemOpen
  \bibfield  {author} {\bibinfo {author} {\bibfnamefont {Walter}\ \bibnamefont
  {Kob}}\ and\ \bibinfo {author} {\bibfnamefont {Hans~C}\ \bibnamefont
  {Andersen}},\ }\bibfield  {title} {\enquote {\bibinfo {title} {Testing
  mode-coupling theory for a supercooled binary lennard-jones mixture i: The
  van hove correlation function},}\ }\href@noop {} {\bibfield  {journal}
  {\bibinfo  {journal} {Physical Review E}\ }\textbf {\bibinfo {volume} {51}},\
  \bibinfo {pages} {4626} (\bibinfo {year} {1995})}\BibitemShut {NoStop}%
\bibitem [{\citenamefont {Kob}(1999)}]{kob1999}%
  \BibitemOpen
  \bibfield  {author} {\bibinfo {author} {\bibfnamefont {Walter}\ \bibnamefont
  {Kob}},\ }\bibfield  {title} {\enquote {\bibinfo {title} {Computer
  simulations of supercooled liquids and glasses},}\ }\href {\doibase
  10.1088/0953-8984/11/10/003} {\bibfield  {journal} {\bibinfo  {journal}
  {Journal of Physics: Condensed Matter}\ }\textbf {\bibinfo {volume} {11}},\
  \bibinfo {pages} {R85} (\bibinfo {year} {1999})}\BibitemShut {NoStop}%
\bibitem [{\citenamefont {Garrahan}\ \emph {et~al.}()\citenamefont {Garrahan},
  \citenamefont {Sollich},\ and\ \citenamefont
  {Toninelli}}]{garrahan2011review}%
  \BibitemOpen
  \bibfield  {author} {\bibinfo {author} {\bibfnamefont {Juan~P}\ \bibnamefont
  {Garrahan}}, \bibinfo {author} {\bibfnamefont {Peter}\ \bibnamefont
  {Sollich}}, \ and\ \bibinfo {author} {\bibfnamefont {Cristina}\ \bibnamefont
  {Toninelli}},\ }\bibfield  {title} {\enquote {\bibinfo {title} {Kinetically
  constrained models},}\ }\href@noop {} {\bibinfo  {journal} {in Dynamical
  Heterogeneities in Glasses, Colloids and Granular Media, edited by L.
  Berthier, G. Biroli, J.-P. Bouchaud, L. Cipelletti, and W. van Saarloosand
  (Oxford University Press, 2011)}\ }\BibitemShut {NoStop}%
\bibitem [{\citenamefont {Scalliet}\ \emph {et~al.}(2022)\citenamefont
  {Scalliet}, \citenamefont {Guiselin},\ and\ \citenamefont
  {Berthier}}]{scalliet2022}%
  \BibitemOpen
\bibfield  {journal} {  }\bibfield  {author} {\bibinfo {author} {\bibfnamefont
  {Camille}\ \bibnamefont {Scalliet}}, \bibinfo {author} {\bibfnamefont
  {Benjamin}\ \bibnamefont {Guiselin}}, \ and\ \bibinfo {author} {\bibfnamefont
  {Ludovic}\ \bibnamefont {Berthier}},\ }\bibfield  {title} {\enquote {\bibinfo
  {title} {Thirty milliseconds in the life of a supercooled liquid},}\ }\href
  {\doibase 10.1103/PhysRevX.12.041028} {\bibfield  {journal} {\bibinfo
  {journal} {Phys. Rev. X}\ }\textbf {\bibinfo {volume} {12}},\ \bibinfo
  {pages} {041028} (\bibinfo {year} {2022})}\BibitemShut {NoStop}%
\bibitem [{\citenamefont {Huang}(1987)}]{kersonhuang1987}%
  \BibitemOpen
  \bibfield  {author} {\bibinfo {author} {\bibfnamefont {Kerson}\ \bibnamefont
  {Huang}},\ }\href@noop {} {\emph {\bibinfo {title} {Statistical Mechanics,
  Second Edition}}}\ (\bibinfo  {publisher} {Wiley},\ \bibinfo {year}
  {1987})\BibitemShut {NoStop}%
\bibitem [{\citenamefont {Fredrickson}\ and\ \citenamefont
  {Andersen}(1984)}]{fredrickson1984}%
  \BibitemOpen
  \bibfield  {author} {\bibinfo {author} {\bibfnamefont {Glenn~H}\ \bibnamefont
  {Fredrickson}}\ and\ \bibinfo {author} {\bibfnamefont {Hans~C}\ \bibnamefont
  {Andersen}},\ }\bibfield  {title} {\enquote {\bibinfo {title} {Kinetic ising
  model of the glass transition},}\ }\href@noop {} {\bibfield  {journal}
  {\bibinfo  {journal} {Physical review letters}\ }\textbf {\bibinfo {volume}
  {53}},\ \bibinfo {pages} {1244} (\bibinfo {year} {1984})}\BibitemShut
  {NoStop}%
\bibitem [{\citenamefont {Biroli}\ and\ \citenamefont
  {M\'ezard}(2001)}]{biroli2001}%
  \BibitemOpen
  \bibfield  {author} {\bibinfo {author} {\bibfnamefont {Giulio}\ \bibnamefont
  {Biroli}}\ and\ \bibinfo {author} {\bibfnamefont {Marc}\ \bibnamefont
  {M\'ezard}},\ }\bibfield  {title} {\enquote {\bibinfo {title} {Lattice glass
  models},}\ }\href {\doibase 10.1103/PhysRevLett.88.025501} {\bibfield
  {journal} {\bibinfo  {journal} {Phys. Rev. Lett.}\ }\textbf {\bibinfo
  {volume} {88}},\ \bibinfo {pages} {025501} (\bibinfo {year}
  {2001})}\BibitemShut {NoStop}%
\bibitem [{\citenamefont {Pica~Ciamarra}\ \emph {et~al.}(2003)\citenamefont
  {Pica~Ciamarra}, \citenamefont {Tarzia}, \citenamefont {de~Candia},\ and\
  \citenamefont {Coniglio}}]{ciamarra2003}%
  \BibitemOpen
  \bibfield  {author} {\bibinfo {author} {\bibfnamefont {M.}~\bibnamefont
  {Pica~Ciamarra}}, \bibinfo {author} {\bibfnamefont {M.}~\bibnamefont
  {Tarzia}}, \bibinfo {author} {\bibfnamefont {A.}~\bibnamefont {de~Candia}}, \
  and\ \bibinfo {author} {\bibfnamefont {A.}~\bibnamefont {Coniglio}},\
  }\bibfield  {title} {\enquote {\bibinfo {title} {Monodisperse model suitable
  to study the glass transition},}\ }\href {\doibase
  10.1103/PhysRevE.68.066111} {\bibfield  {journal} {\bibinfo  {journal} {Phys.
  Rev. E}\ }\textbf {\bibinfo {volume} {68}},\ \bibinfo {pages} {066111}
  (\bibinfo {year} {2003})}\BibitemShut {NoStop}%
\bibitem [{\citenamefont {Krzakala}\ \emph {et~al.}(2008)\citenamefont
  {Krzakala}, \citenamefont {Tarzia},\ and\ \citenamefont
  {Zdeborov\'a}}]{krzakala2008}%
  \BibitemOpen
  \bibfield  {author} {\bibinfo {author} {\bibfnamefont {Florent}\ \bibnamefont
  {Krzakala}}, \bibinfo {author} {\bibfnamefont {Marco}\ \bibnamefont
  {Tarzia}}, \ and\ \bibinfo {author} {\bibfnamefont {Lenka}\ \bibnamefont
  {Zdeborov\'a}},\ }\bibfield  {title} {\enquote {\bibinfo {title} {Lattice
  model for colloidal gels and glasses},}\ }\href {\doibase
  10.1103/PhysRevLett.101.165702} {\bibfield  {journal} {\bibinfo  {journal}
  {Phys. Rev. Lett.}\ }\textbf {\bibinfo {volume} {101}},\ \bibinfo {pages}
  {165702} (\bibinfo {year} {2008})}\BibitemShut {NoStop}%
\bibitem [{\citenamefont {Nishikawa}\ and\ \citenamefont
  {Hukushima}(2020)}]{nishikawa2020}%
  \BibitemOpen
  \bibfield  {author} {\bibinfo {author} {\bibfnamefont {Yoshihiko}\
  \bibnamefont {Nishikawa}}\ and\ \bibinfo {author} {\bibfnamefont {Koji}\
  \bibnamefont {Hukushima}},\ }\bibfield  {title} {\enquote {\bibinfo {title}
  {Lattice glass model in three spatial dimensions},}\ }\href {\doibase
  10.1103/PhysRevLett.125.065501} {\bibfield  {journal} {\bibinfo  {journal}
  {Phys. Rev. Lett.}\ }\textbf {\bibinfo {volume} {125}},\ \bibinfo {pages}
  {065501} (\bibinfo {year} {2020})}\BibitemShut {NoStop}%
\bibitem [{\citenamefont {Zhang}\ and\ \citenamefont {Lam}(2017)}]{zhang2017}%
  \BibitemOpen
  \bibfield  {author} {\bibinfo {author} {\bibfnamefont {Ling-Han}\
  \bibnamefont {Zhang}}\ and\ \bibinfo {author} {\bibfnamefont {Chi-Hang}\
  \bibnamefont {Lam}},\ }\bibfield  {title} {\enquote {\bibinfo {title}
  {Emergent facilitation behavior in a distinguishable-particle lattice model
  of glass},}\ }\href {\doibase 10.1103/PhysRevB.95.184202} {\bibfield
  {journal} {\bibinfo  {journal} {Phys. Rev. B}\ }\textbf {\bibinfo {volume}
  {95}},\ \bibinfo {pages} {184202} (\bibinfo {year} {2017})}\BibitemShut
  {NoStop}%
\bibitem [{\citenamefont {Hasyim}\ and\ \citenamefont
  {Mandadapu}(2023)}]{hasyim2023}%
  \BibitemOpen
  \bibfield  {author} {\bibinfo {author} {\bibfnamefont {Muhammad~R.}\
  \bibnamefont {Hasyim}}\ and\ \bibinfo {author} {\bibfnamefont {Kranthi~K.}\
  \bibnamefont {Mandadapu}},\ }\bibfield  {title} {\enquote {\bibinfo {title}
  {Emergent facilitation and glassy dynamics in supercooled liquids},}\ }\href
  {\doibase 10.48550/ARXIV.2310.06584} {\bibfield  {journal} {\bibinfo
  {journal} {arXiv}\ } (\bibinfo {year} {2023}),\
  10.48550/ARXIV.2310.06584}\BibitemShut {NoStop}%
\bibitem [{\citenamefont {Lulli}\ \emph {et~al.}(2020)\citenamefont {Lulli},
  \citenamefont {Lee}, \citenamefont {Deng}, \citenamefont {Yip},\ and\
  \citenamefont {Lam}}]{lulli2020}%
  \BibitemOpen
  \bibfield  {author} {\bibinfo {author} {\bibfnamefont {Matteo}\ \bibnamefont
  {Lulli}}, \bibinfo {author} {\bibfnamefont {Chun-Shing}\ \bibnamefont {Lee}},
  \bibinfo {author} {\bibfnamefont {Hai-Yao}\ \bibnamefont {Deng}}, \bibinfo
  {author} {\bibfnamefont {Cho-Tung}\ \bibnamefont {Yip}}, \ and\ \bibinfo
  {author} {\bibfnamefont {Chi-Hang}\ \bibnamefont {Lam}},\ }\bibfield  {title}
  {\enquote {\bibinfo {title} {Spatial heterogeneities in structural
  temperature cause kovacs' expansion gap paradox in aging of glasses},}\
  }\href@noop {} {\bibfield  {journal} {\bibinfo  {journal} {Physical Review
  Letters}\ }\textbf {\bibinfo {volume} {124}},\ \bibinfo {pages} {095501}
  (\bibinfo {year} {2020})}\BibitemShut {NoStop}%
\bibitem [{\citenamefont {Lee}\ \emph {et~al.}(2020)\citenamefont {Lee},
  \citenamefont {Lulli}, \citenamefont {Zhang}, \citenamefont {Deng},\ and\
  \citenamefont {Lam}}]{lee2020}%
  \BibitemOpen
  \bibfield  {author} {\bibinfo {author} {\bibfnamefont {Chun-Shing}\
  \bibnamefont {Lee}}, \bibinfo {author} {\bibfnamefont {Matteo}\ \bibnamefont
  {Lulli}}, \bibinfo {author} {\bibfnamefont {Ling-Han}\ \bibnamefont {Zhang}},
  \bibinfo {author} {\bibfnamefont {Hai-Yao}\ \bibnamefont {Deng}}, \ and\
  \bibinfo {author} {\bibfnamefont {Chi-Hang}\ \bibnamefont {Lam}},\ }\bibfield
   {title} {\enquote {\bibinfo {title} {Fragile glasses associated with a
  dramatic drop of entropy under supercooling},}\ }\href@noop {} {\bibfield
  {journal} {\bibinfo  {journal} {Physical Review Letters}\ }\textbf {\bibinfo
  {volume} {125}},\ \bibinfo {pages} {265703} (\bibinfo {year}
  {2020})}\BibitemShut {NoStop}%
\bibitem [{\citenamefont {Yip}\ \emph {et~al.}(2020)\citenamefont {Yip},
  \citenamefont {Isobe}, \citenamefont {Chan}, \citenamefont {Ren},
  \citenamefont {Wong}, \citenamefont {Huo}, \citenamefont {Lee}, \citenamefont
  {Tsang}, \citenamefont {Han},\ and\ \citenamefont {Lam}}]{yip2020}%
  \BibitemOpen
  \bibfield  {author} {\bibinfo {author} {\bibfnamefont {Cho-Tung}\
  \bibnamefont {Yip}}, \bibinfo {author} {\bibfnamefont {Masaharu}\
  \bibnamefont {Isobe}}, \bibinfo {author} {\bibfnamefont {Chor-Hoi}\
  \bibnamefont {Chan}}, \bibinfo {author} {\bibfnamefont {Simiao}\ \bibnamefont
  {Ren}}, \bibinfo {author} {\bibfnamefont {Kin-Ping}\ \bibnamefont {Wong}},
  \bibinfo {author} {\bibfnamefont {Qingxiao}\ \bibnamefont {Huo}}, \bibinfo
  {author} {\bibfnamefont {Chun-Sing}\ \bibnamefont {Lee}}, \bibinfo {author}
  {\bibfnamefont {Yuen-Hong}\ \bibnamefont {Tsang}}, \bibinfo {author}
  {\bibfnamefont {Yilong}\ \bibnamefont {Han}}, \ and\ \bibinfo {author}
  {\bibfnamefont {Chi-Hang}\ \bibnamefont {Lam}},\ }\bibfield  {title}
  {\enquote {\bibinfo {title} {Direct evidence of void-induced structural
  relaxations in colloidal glass formers},}\ }\href@noop {} {\bibfield
  {journal} {\bibinfo  {journal} {Physical Review Letters}\ }\textbf {\bibinfo
  {volume} {125}},\ \bibinfo {pages} {258001} (\bibinfo {year}
  {2020})}\BibitemShut {NoStop}%
\bibitem [{\citenamefont {Lee}\ \emph {et~al.}(2021)\citenamefont {Lee},
  \citenamefont {Deng}, \citenamefont {Yip},\ and\ \citenamefont
  {Lam}}]{lee2021}%
  \BibitemOpen
  \bibfield  {author} {\bibinfo {author} {\bibfnamefont {Chun-Shing}\
  \bibnamefont {Lee}}, \bibinfo {author} {\bibfnamefont {Hai-Yao}\ \bibnamefont
  {Deng}}, \bibinfo {author} {\bibfnamefont {Cho-Tung}\ \bibnamefont {Yip}}, \
  and\ \bibinfo {author} {\bibfnamefont {Chi-Hang}\ \bibnamefont {Lam}},\
  }\bibfield  {title} {\enquote {\bibinfo {title} {Large heat-capacity jump in
  cooling-heating of fragile glass from kinetic monte carlo simulations based
  on a two-state picture},}\ }\href@noop {} {\bibfield  {journal} {\bibinfo
  {journal} {Physical Review E}\ }\textbf {\bibinfo {volume} {104}},\ \bibinfo
  {pages} {024131} (\bibinfo {year} {2021})}\BibitemShut {NoStop}%
\bibitem [{\citenamefont {Gao}\ \emph {et~al.}(2022)\citenamefont {Gao},
  \citenamefont {Deng}, \citenamefont {Lee}, \citenamefont {You},\ and\
  \citenamefont {Lam}}]{gao2022}%
  \BibitemOpen
  \bibfield  {author} {\bibinfo {author} {\bibfnamefont {Xin-Yuan}\
  \bibnamefont {Gao}}, \bibinfo {author} {\bibfnamefont {Hai-Yao}\ \bibnamefont
  {Deng}}, \bibinfo {author} {\bibfnamefont {Chun-Shing}\ \bibnamefont {Lee}},
  \bibinfo {author} {\bibfnamefont {Jian-Qiang}\ \bibnamefont {You}}, \ and\
  \bibinfo {author} {\bibfnamefont {Chi-Hang}\ \bibnamefont {Lam}},\ }\bibfield
   {title} {\enquote {\bibinfo {title} {Emergence of two-level systems in glass
  formers: a kinetic monte carlo study},}\ }\href@noop {} {\bibfield  {journal}
  {\bibinfo  {journal} {Soft Matter}\ }\textbf {\bibinfo {volume} {18}},\
  \bibinfo {pages} {2211--2221} (\bibinfo {year} {2022})}\BibitemShut {NoStop}%
\bibitem [{\citenamefont {Gopinath}\ \emph {et~al.}(2022)\citenamefont
  {Gopinath}, \citenamefont {Lee}, \citenamefont {Gao}, \citenamefont {An},
  \citenamefont {Chan}, \citenamefont {Yip}, \citenamefont {Deng},\ and\
  \citenamefont {Lam}}]{gopinath2022}%
  \BibitemOpen
  \bibfield  {author} {\bibinfo {author} {\bibfnamefont {Gautham}\ \bibnamefont
  {Gopinath}}, \bibinfo {author} {\bibfnamefont {Chun-Shing}\ \bibnamefont
  {Lee}}, \bibinfo {author} {\bibfnamefont {Xin-Yuan}\ \bibnamefont {Gao}},
  \bibinfo {author} {\bibfnamefont {Xiao-Dong}\ \bibnamefont {An}}, \bibinfo
  {author} {\bibfnamefont {Chor-Hoi}\ \bibnamefont {Chan}}, \bibinfo {author}
  {\bibfnamefont {Cho-Tung}\ \bibnamefont {Yip}}, \bibinfo {author}
  {\bibfnamefont {Hai-Yao}\ \bibnamefont {Deng}}, \ and\ \bibinfo {author}
  {\bibfnamefont {Chi-Hang}\ \bibnamefont {Lam}},\ }\bibfield  {title}
  {\enquote {\bibinfo {title} {Diffusion-coefficient power laws and
  defect-driven glassy dynamics in swap acceleration},}\ }\href@noop {}
  {\bibfield  {journal} {\bibinfo  {journal} {Physical Review Letters}\
  }\textbf {\bibinfo {volume} {129}},\ \bibinfo {pages} {168002} (\bibinfo
  {year} {2022})}\BibitemShut {NoStop}%
\bibitem [{\citenamefont {Gao}\ \emph {et~al.}(2023)\citenamefont {Gao},
  \citenamefont {Ong}, \citenamefont {Lee}, \citenamefont {Yip}, \citenamefont
  {Deng},\ and\ \citenamefont {Lam}}]{gao2023Kauzmann}%
  \BibitemOpen
  \bibfield  {author} {\bibinfo {author} {\bibfnamefont {Xin-Yuan}\
  \bibnamefont {Gao}}, \bibinfo {author} {\bibfnamefont {Chin-Yuan}\
  \bibnamefont {Ong}}, \bibinfo {author} {\bibfnamefont {Chun-Shing}\
  \bibnamefont {Lee}}, \bibinfo {author} {\bibfnamefont {Cho-Tung}\
  \bibnamefont {Yip}}, \bibinfo {author} {\bibfnamefont {Hai-Yao}\ \bibnamefont
  {Deng}}, \ and\ \bibinfo {author} {\bibfnamefont {Chi-Hang}\ \bibnamefont
  {Lam}},\ }\bibfield  {title} {\enquote {\bibinfo {title} {Kauzmann paradox: A
  possible crossover due to diminishing local excitations},}\ }\href {\doibase
  10.1103/PhysRevB.107.174206} {\bibfield  {journal} {\bibinfo  {journal}
  {Phys. Rev. B}\ }\textbf {\bibinfo {volume} {107}},\ \bibinfo {pages}
  {174206} (\bibinfo {year} {2023})}\BibitemShut {NoStop}%
\bibitem [{\citenamefont {Zinn-Justin}(2002)}]{justin2002}%
  \BibitemOpen
  \bibfield  {author} {\bibinfo {author} {\bibfnamefont {J.}~\bibnamefont
  {Zinn-Justin}},\ }\href@noop {} {\emph {\bibinfo {title} {Quantum Field
  Theory and Critical Phenomena}}},\ International series of monographs on
  physics\ (\bibinfo  {publisher} {Clarendon Press},\ \bibinfo {year}
  {2002})\BibitemShut {NoStop}%
\bibitem [{\citenamefont {Flenner}\ and\ \citenamefont
  {Szamel}(2015)}]{flenner2015}%
  \BibitemOpen
  \bibfield  {author} {\bibinfo {author} {\bibfnamefont {Elijah}\ \bibnamefont
  {Flenner}}\ and\ \bibinfo {author} {\bibfnamefont {Grzegorz}\ \bibnamefont
  {Szamel}},\ }\bibfield  {title} {\enquote {\bibinfo {title} {Fundamental
  differences between glassy dynamics in two and three dimensions},}\
  }\href@noop {} {\bibfield  {journal} {\bibinfo  {journal} {Nat. Commun.}\
  }\textbf {\bibinfo {volume} {6}},\ \bibinfo {pages} {7392} (\bibinfo {year}
  {2015})}\BibitemShut {NoStop}%
\bibitem [{\citenamefont {Shiba}\ \emph {et~al.}(2016)\citenamefont {Shiba},
  \citenamefont {Yamada}, \citenamefont {Kawasaki},\ and\ \citenamefont
  {Kim}}]{shiba2016}%
  \BibitemOpen
  \bibfield  {author} {\bibinfo {author} {\bibfnamefont {H.}~\bibnamefont
  {Shiba}}, \bibinfo {author} {\bibfnamefont {Y.}~\bibnamefont {Yamada}},
  \bibinfo {author} {\bibfnamefont {T.}~\bibnamefont {Kawasaki}}, \ and\
  \bibinfo {author} {\bibfnamefont {K.}~\bibnamefont {Kim}},\ }\bibfield
  {title} {\enquote {\bibinfo {title} {Unveiling dimensionality dependence of
  glassy dynamics: 2d infinite fluctuation eclipses inherent structural
  relaxation},}\ }\href@noop {} {\bibfield  {journal} {\bibinfo  {journal}
  {Phys. Rev. Lett.}\ }\textbf {\bibinfo {volume} {117}},\ \bibinfo {pages}
  {245701} (\bibinfo {year} {2016})}\BibitemShut {NoStop}%
\bibitem [{\citenamefont {Vivek}\ \emph {et~al.}(2017)\citenamefont {Vivek},
  \citenamefont {Kelleher}, \citenamefont {Chaikin},\ and\ \citenamefont
  {Weeks}}]{weeks2017}%
  \BibitemOpen
  \bibfield  {author} {\bibinfo {author} {\bibfnamefont {Skanda}\ \bibnamefont
  {Vivek}}, \bibinfo {author} {\bibfnamefont {Colm~P}\ \bibnamefont
  {Kelleher}}, \bibinfo {author} {\bibfnamefont {Paul~M}\ \bibnamefont
  {Chaikin}}, \ and\ \bibinfo {author} {\bibfnamefont {Eric~R}\ \bibnamefont
  {Weeks}},\ }\bibfield  {title} {\enquote {\bibinfo {title} {Long-wavelength
  fluctuations and the glass transition in two dimensions and three
  dimensions},}\ }\href@noop {} {\bibfield  {journal} {\bibinfo  {journal}
  {Proceedings of the National Academy of Sciences}\ }\textbf {\bibinfo
  {volume} {114}},\ \bibinfo {pages} {1850--1855} (\bibinfo {year}
  {2017})}\BibitemShut {NoStop}%
\bibitem [{\citenamefont {Flenner}\ and\ \citenamefont
  {Szamel}(2019)}]{Flenner2019}%
  \BibitemOpen
  \bibfield  {author} {\bibinfo {author} {\bibfnamefont {Elijah}\ \bibnamefont
  {Flenner}}\ and\ \bibinfo {author} {\bibfnamefont {Grzegorz}\ \bibnamefont
  {Szamel}},\ }\bibfield  {title} {\enquote {\bibinfo {title} {Viscoelastic
  shear stress relaxation in two-dimensional glass-forming liquids},}\ }\href
  {\doibase 10.1073/pnas.1815097116} {\bibfield  {journal} {\bibinfo  {journal}
  {Proceedings of the National Academy of Sciences}\ }\textbf {\bibinfo
  {volume} {116}},\ \bibinfo {pages} {2015--2020} (\bibinfo {year}
  {2019})}\BibitemShut {NoStop}%
\bibitem [{\citenamefont {Berthier}\ \emph {et~al.}(2019)\citenamefont
  {Berthier}, \citenamefont {Charbonneau}, \citenamefont {Ninarello},
  \citenamefont {Ozawa},\ and\ \citenamefont {Yaida}}]{berthier2019zero}%
  \BibitemOpen
  \bibfield  {author} {\bibinfo {author} {\bibfnamefont {Ludovic}\ \bibnamefont
  {Berthier}}, \bibinfo {author} {\bibfnamefont {Patrick}\ \bibnamefont
  {Charbonneau}}, \bibinfo {author} {\bibfnamefont {Andrea}\ \bibnamefont
  {Ninarello}}, \bibinfo {author} {\bibfnamefont {Misaki}\ \bibnamefont
  {Ozawa}}, \ and\ \bibinfo {author} {\bibfnamefont {Sho}\ \bibnamefont
  {Yaida}},\ }\bibfield  {title} {\enquote {\bibinfo {title} {Zero-temperature
  glass transition in two dimensions},}\ }\href@noop {} {\bibfield  {journal}
  {\bibinfo  {journal} {Nature communications}\ }\textbf {\bibinfo {volume}
  {10}},\ \bibinfo {pages} {1--7} (\bibinfo {year} {2019})}\BibitemShut
  {NoStop}%
\bibitem [{\citenamefont {Newman}\ and\ \citenamefont
  {Moore}(1999)}]{newman1999}%
  \BibitemOpen
  \bibfield  {author} {\bibinfo {author} {\bibfnamefont {MEJ}\ \bibnamefont
  {Newman}}\ and\ \bibinfo {author} {\bibfnamefont {Cristopher}\ \bibnamefont
  {Moore}},\ }\bibfield  {title} {\enquote {\bibinfo {title} {Glassy dynamics
  and aging in an exactly solvable spin model},}\ }\href@noop {} {\bibfield
  {journal} {\bibinfo  {journal} {Physical Review E}\ }\textbf {\bibinfo
  {volume} {60}},\ \bibinfo {pages} {5068} (\bibinfo {year}
  {1999})}\BibitemShut {NoStop}%
\bibitem [{\citenamefont {Jack}\ and\ \citenamefont
  {Garrahan}(2016)}]{jack2016}%
  \BibitemOpen
  \bibfield  {author} {\bibinfo {author} {\bibfnamefont {Robert~L}\
  \bibnamefont {Jack}}\ and\ \bibinfo {author} {\bibfnamefont {Juan~P}\
  \bibnamefont {Garrahan}},\ }\bibfield  {title} {\enquote {\bibinfo {title}
  {Phase transition for quenched coupled replicas in a plaquette spin model of
  glasses},}\ }\href@noop {} {\bibfield  {journal} {\bibinfo  {journal}
  {Physical review letters}\ }\textbf {\bibinfo {volume} {116}},\ \bibinfo
  {pages} {055702} (\bibinfo {year} {2016})}\BibitemShut {NoStop}%
\bibitem [{\citenamefont {Ritort}\ and\ \citenamefont
  {Sollich}(2003)}]{ritort2003review}%
  \BibitemOpen
  \bibfield  {author} {\bibinfo {author} {\bibfnamefont {Felix}\ \bibnamefont
  {Ritort}}\ and\ \bibinfo {author} {\bibfnamefont {Peter}\ \bibnamefont
  {Sollich}},\ }\bibfield  {title} {\enquote {\bibinfo {title} {Glassy dynamics
  of kinetically constrained models},}\ }\href@noop {} {\bibfield  {journal}
  {\bibinfo  {journal} {Advances in Physics}\ }\textbf {\bibinfo {volume}
  {52}},\ \bibinfo {pages} {219--342} (\bibinfo {year} {2003})}\BibitemShut
  {NoStop}%
\bibitem [{\citenamefont {G{\H o}tze}(2008)}]{Gotzebook}%
  \BibitemOpen
  \bibfield  {author} {\bibinfo {author} {\bibfnamefont {Wolfgang}\
  \bibnamefont {G{\H o}tze}},\ }\href@noop {} {\emph {\bibinfo {title}
  {{Complex dynamics of glass-forming liquids: a mode-coupling theory}}}}\
  (\bibinfo  {publisher} {Oxford University Press},\ \bibinfo {year}
  {2008})\BibitemShut {NoStop}%
\bibitem [{\citenamefont {Lam}(2018)}]{lam2018tree}%
  \BibitemOpen
  \bibfield  {author} {\bibinfo {author} {\bibfnamefont {Chi-Hang}\
  \bibnamefont {Lam}},\ }\bibfield  {title} {\enquote {\bibinfo {title} {Local
  random configuration-tree theory for string repetition and facilitated
  dynamics of glass},}\ }\href@noop {} {\bibfield  {journal} {\bibinfo
  {journal} {Journal of Statistical Mechanics: Theory and Experiment}\ }\textbf
  {\bibinfo {volume} {2018}},\ \bibinfo {pages} {023301} (\bibinfo {year}
  {2018})}\BibitemShut {NoStop}%
\bibitem [{\citenamefont {Deng}\ \emph {et~al.}(2019)\citenamefont {Deng},
  \citenamefont {Lee}, \citenamefont {Lulli}, \citenamefont {Zhang},\ and\
  \citenamefont {Lam}}]{deng2019}%
  \BibitemOpen
  \bibfield  {author} {\bibinfo {author} {\bibfnamefont {Hai-Yao}\ \bibnamefont
  {Deng}}, \bibinfo {author} {\bibfnamefont {Chun-Shing}\ \bibnamefont {Lee}},
  \bibinfo {author} {\bibfnamefont {Matteo}\ \bibnamefont {Lulli}}, \bibinfo
  {author} {\bibfnamefont {Ling-Han}\ \bibnamefont {Zhang}}, \ and\ \bibinfo
  {author} {\bibfnamefont {Chi-Hang}\ \bibnamefont {Lam}},\ }\bibfield  {title}
  {\enquote {\bibinfo {title} {Configuration-tree theoretical calculation of
  the mean-squared displacement of particles in glass formers},}\ }\href
  {\doibase 10.1088/1742-5468/ab39d7} {\bibfield  {journal} {\bibinfo
  {journal} {Journal of Statistical Mechanics: Theory and Experiment}\ }\textbf
  {\bibinfo {volume} {2019}},\ \bibinfo {pages} {094014} (\bibinfo {year}
  {2019})}\BibitemShut {NoStop}%
\bibitem [{\citenamefont {Moynihan}\ and\ \citenamefont
  {Angell}(2000)}]{angell2000}%
  \BibitemOpen
  \bibfield  {author} {\bibinfo {author} {\bibfnamefont {Cornelius~T}\
  \bibnamefont {Moynihan}}\ and\ \bibinfo {author} {\bibfnamefont {C~Austen}\
  \bibnamefont {Angell}},\ }\bibfield  {title} {\enquote {\bibinfo {title}
  {Bond lattice or excitation model analysis of the configurational entropy of
  molecular liquids},}\ }\href@noop {} {\bibfield  {journal} {\bibinfo
  {journal} {Journal of non-crystalline solids}\ }\textbf {\bibinfo {volume}
  {274}},\ \bibinfo {pages} {131--138} (\bibinfo {year} {2000})}\BibitemShut
  {NoStop}%
\bibitem [{\citenamefont {Ong}\ \emph {et~al.}(2022)\citenamefont {Ong},
  \citenamefont {Lee}, \citenamefont {Gao}, \citenamefont {Zhai}, \citenamefont
  {Shi}, \citenamefont {Deng},\ and\ \citenamefont {Lam}}]{ong2022}%
  \BibitemOpen
  \bibfield  {author} {\bibinfo {author} {\bibfnamefont {Chin-Yuan}\
  \bibnamefont {Ong}}, \bibinfo {author} {\bibfnamefont {Chun-Shing}\
  \bibnamefont {Lee}}, \bibinfo {author} {\bibfnamefont {Xin-Yuan}\
  \bibnamefont {Gao}}, \bibinfo {author} {\bibfnamefont {Qiang}\ \bibnamefont
  {Zhai}}, \bibinfo {author} {\bibfnamefont {Rui}\ \bibnamefont {Shi}},
  \bibinfo {author} {\bibfnamefont {Hai-Yao}\ \bibnamefont {Deng}}, \ and\
  \bibinfo {author} {\bibfnamefont {Chi-Hang}\ \bibnamefont {Lam}},\ }\bibfield
   {title} {\enquote {\bibinfo {title} {Towards relating fragile-to-strong
  transition to fragile glass},}\ }\href {\doibase 10.48550/ARXIV.2211.15026}
  {\bibfield  {journal} {\bibinfo  {journal} {arXiv:2211.15026}\ } (\bibinfo
  {year} {2022}),\ 10.48550/ARXIV.2211.15026}\BibitemShut {NoStop}%
\bibitem [{\citenamefont {Dawson}\ \emph {et~al.}(2000)\citenamefont {Dawson},
  \citenamefont {Foffi}, \citenamefont {Fuchs}, \citenamefont {G\"otze},
  \citenamefont {Sciortino}, \citenamefont {Sperl}, \citenamefont {Tartaglia},
  \citenamefont {Voigtmann},\ and\ \citenamefont {Zaccarelli}}]{dsawson2000}%
  \BibitemOpen
  \bibfield  {author} {\bibinfo {author} {\bibfnamefont {K.}~\bibnamefont
  {Dawson}}, \bibinfo {author} {\bibfnamefont {G.}~\bibnamefont {Foffi}},
  \bibinfo {author} {\bibfnamefont {M.}~\bibnamefont {Fuchs}}, \bibinfo
  {author} {\bibfnamefont {W.}~\bibnamefont {G\"otze}}, \bibinfo {author}
  {\bibfnamefont {F.}~\bibnamefont {Sciortino}}, \bibinfo {author}
  {\bibfnamefont {M.}~\bibnamefont {Sperl}}, \bibinfo {author} {\bibfnamefont
  {P.}~\bibnamefont {Tartaglia}}, \bibinfo {author} {\bibfnamefont {Th.}\
  \bibnamefont {Voigtmann}}, \ and\ \bibinfo {author} {\bibfnamefont
  {E.}~\bibnamefont {Zaccarelli}},\ }\bibfield  {title} {\enquote {\bibinfo
  {title} {Higher-order glass-transition singularities in colloidal systems
  with attractive interactions},}\ }\href {\doibase 10.1103/PhysRevE.63.011401}
  {\bibfield  {journal} {\bibinfo  {journal} {Phys. Rev. E}\ }\textbf {\bibinfo
  {volume} {63}},\ \bibinfo {pages} {011401} (\bibinfo {year}
  {2000})}\BibitemShut {NoStop}%
\bibitem [{\citenamefont {Saika-Voivod}\ \emph {et~al.}(2001)\citenamefont
  {Saika-Voivod}, \citenamefont {Poole},\ and\ \citenamefont
  {Sciortino}}]{SaikaVoivod2001}%
  \BibitemOpen
  \bibfield  {author} {\bibinfo {author} {\bibfnamefont {Ivan}\ \bibnamefont
  {Saika-Voivod}}, \bibinfo {author} {\bibfnamefont {Peter~H.}\ \bibnamefont
  {Poole}}, \ and\ \bibinfo {author} {\bibfnamefont {Francesco}\ \bibnamefont
  {Sciortino}},\ }\bibfield  {title} {\enquote {\bibinfo {title}
  {Fragile-to-strong transition and polyamorphism in the energy landscape of
  liquid silica},}\ }\href {\doibase 10.1038/35087524} {\bibfield  {journal}
  {\bibinfo  {journal} {Nature}\ }\textbf {\bibinfo {volume} {412}},\ \bibinfo
  {pages} {514--517} (\bibinfo {year} {2001})}\BibitemShut {NoStop}%
\bibitem [{\citenamefont {Ciarella}\ \emph {et~al.}(2019)\citenamefont
  {Ciarella}, \citenamefont {Biezemans},\ and\ \citenamefont
  {Janssen}}]{ciarella2019pnas}%
  \BibitemOpen
  \bibfield  {author} {\bibinfo {author} {\bibfnamefont {Simone}\ \bibnamefont
  {Ciarella}}, \bibinfo {author} {\bibfnamefont {Rutger~A.}\ \bibnamefont
  {Biezemans}}, \ and\ \bibinfo {author} {\bibfnamefont {Liesbeth M.~C.}\
  \bibnamefont {Janssen}},\ }\bibfield  {title} {\enquote {\bibinfo {title}
  {Understanding, predicting, and tuning the fragility of vitrimeric
  polymers},}\ }\href {\doibase 10.1073/pnas.1912571116} {\bibfield  {journal}
  {\bibinfo  {journal} {Proceedings of the National Academy of Sciences}\
  }\textbf {\bibinfo {volume} {116}},\ \bibinfo {pages} {25013–25022}
  (\bibinfo {year} {2019})}\BibitemShut {NoStop}%
\bibitem [{\citenamefont {Liao}\ and\ \citenamefont {Chen}(2001)}]{liao2001v2}%
  \BibitemOpen
  \bibfield  {author} {\bibinfo {author} {\bibfnamefont {C.~Y.}\ \bibnamefont
  {Liao}}\ and\ \bibinfo {author} {\bibfnamefont {S.-H.}\ \bibnamefont
  {Chen}},\ }\bibfield  {title} {\enquote {\bibinfo {title} {Dynamics of
  inherent structure in supercooled liquids near kinetic glass transition},}\
  }\href {\doibase 10.1103/PhysRevE.64.031202} {\bibfield  {journal} {\bibinfo
  {journal} {Phys. Rev. E}\ }\textbf {\bibinfo {volume} {64}},\ \bibinfo
  {pages} {031202} (\bibinfo {year} {2001})}\BibitemShut {NoStop}%
\bibitem [{\citenamefont {Yamamoto}\ and\ \citenamefont
  {Onuki}(1998)}]{yamamoto1998}%
  \BibitemOpen
  \bibfield  {author} {\bibinfo {author} {\bibfnamefont {Ryoichi}\ \bibnamefont
  {Yamamoto}}\ and\ \bibinfo {author} {\bibfnamefont {Akira}\ \bibnamefont
  {Onuki}},\ }\bibfield  {title} {\enquote {\bibinfo {title} {Dynamics of
  highly supercooled liquids: Heterogeneity, rheology, and diffusion},}\ }\href
  {\doibase 10.1103/PhysRevE.58.3515} {\bibfield  {journal} {\bibinfo
  {journal} {Phys. Rev. E}\ }\textbf {\bibinfo {volume} {58}},\ \bibinfo
  {pages} {3515--3529} (\bibinfo {year} {1998})}\BibitemShut {NoStop}%
\bibitem [{\citenamefont {Kawasaki}\ and\ \citenamefont
  {Kim}(2017)}]{kawasaki2017}%
  \BibitemOpen
  \bibfield  {author} {\bibinfo {author} {\bibfnamefont {Takeshi}\ \bibnamefont
  {Kawasaki}}\ and\ \bibinfo {author} {\bibfnamefont {Kang}\ \bibnamefont
  {Kim}},\ }\bibfield  {title} {\enquote {\bibinfo {title} {Identifying time
  scales for violation/preservation of stokes-einstein relation in supercooled
  water},}\ }\href {\doibase 10.1126/sciadv.1700399} {\bibfield  {journal}
  {\bibinfo  {journal} {Science Advances}\ }\textbf {\bibinfo {volume} {3}},\
  \bibinfo {pages} {1} (\bibinfo {year} {2017})}\BibitemShut {NoStop}%
\bibitem [{\citenamefont {Shiba}\ \emph {et~al.}(2019)\citenamefont {Shiba},
  \citenamefont {Kawasaki},\ and\ \citenamefont {Kim}}]{shiba2019}%
  \BibitemOpen
  \bibfield  {author} {\bibinfo {author} {\bibfnamefont {Hayato}\ \bibnamefont
  {Shiba}}, \bibinfo {author} {\bibfnamefont {Takeshi}\ \bibnamefont
  {Kawasaki}}, \ and\ \bibinfo {author} {\bibfnamefont {Kang}\ \bibnamefont
  {Kim}},\ }\bibfield  {title} {\enquote {\bibinfo {title} {Local density
  fluctuation governs the divergence of viscosity underlying elastic and
  hydrodynamic anomalies in a 2d glass-forming liquid},}\ }\href {\doibase
  10.1103/PhysRevLett.123.265501} {\bibfield  {journal} {\bibinfo  {journal}
  {Phys. Rev. Lett.}\ }\textbf {\bibinfo {volume} {123}},\ \bibinfo {pages}
  {265501} (\bibinfo {year} {2019})}\BibitemShut {NoStop}%
\bibitem [{\citenamefont {Kawasaki}\ and\ \citenamefont
  {Onuki}(2013)}]{kawasaki2013}%
  \BibitemOpen
  \bibfield  {author} {\bibinfo {author} {\bibfnamefont {Takeshi}\ \bibnamefont
  {Kawasaki}}\ and\ \bibinfo {author} {\bibfnamefont {Akira}\ \bibnamefont
  {Onuki}},\ }\bibfield  {title} {\enquote {\bibinfo {title} {Slow relaxations
  and stringlike jump motions in fragile glass-forming liquids: Breakdown of
  the stokes-einstein relation},}\ }\href@noop {} {\bibfield  {journal}
  {\bibinfo  {journal} {Phys. Rev. E}\ }\textbf {\bibinfo {volume} {87}},\
  \bibinfo {pages} {012312} (\bibinfo {year} {2013})}\BibitemShut {NoStop}%
\bibitem [{\citenamefont {Berthier}\ \emph
  {et~al.}(2011{\natexlab{a}})\citenamefont {Berthier}, \citenamefont {Biroli},
  \citenamefont {Bouchaud}, \citenamefont {Cipelletti},\ and\ \citenamefont
  {van Saarloos}}]{berthier2011book}%
  \BibitemOpen
  \bibfield  {author} {\bibinfo {author} {\bibfnamefont {Ludovic}\ \bibnamefont
  {Berthier}}, \bibinfo {author} {\bibfnamefont {Giulio}\ \bibnamefont
  {Biroli}}, \bibinfo {author} {\bibfnamefont {Jean-Philippe}\ \bibnamefont
  {Bouchaud}}, \bibinfo {author} {\bibfnamefont {Luca}\ \bibnamefont
  {Cipelletti}}, \ and\ \bibinfo {author} {\bibfnamefont {Wim}\ \bibnamefont
  {van Saarloos}},\ }\href@noop {} {\emph {\bibinfo {title} {Dynamical
  heterogeneities in glasses, colloids, and granular media}}},\ Vol.\ \bibinfo
  {volume} {150}\ (\bibinfo  {publisher} {Oxford University Press},\ \bibinfo
  {year} {2011})\BibitemShut {NoStop}%
\bibitem [{\citenamefont {Berthier}\ \emph
  {et~al.}(2011{\natexlab{b}})\citenamefont {Berthier}, \citenamefont {Biroli},
  \citenamefont {Bouchaud},\ and\ \citenamefont {Jack}}]{berthier2011reviewv2}%
  \BibitemOpen
  \bibfield  {author} {\bibinfo {author} {\bibfnamefont {Ludovic}\ \bibnamefont
  {Berthier}}, \bibinfo {author} {\bibfnamefont {Giulio}\ \bibnamefont
  {Biroli}}, \bibinfo {author} {\bibfnamefont {Jean-Philippe}\ \bibnamefont
  {Bouchaud}}, \ and\ \bibinfo {author} {\bibfnamefont {Robert~L}\ \bibnamefont
  {Jack}},\ }\bibfield  {title} {\enquote {\bibinfo {title} {Overview of
  different characterisations of dynamic heterogeneity},}\ }\href@noop {}
  {\bibfield  {journal} {\bibinfo  {journal} {Dynamical Heterogeneities in
  Glasses, Colloids, and Granular Media}\ }\textbf {\bibinfo {volume} {150}},\
  \bibinfo {pages} {68} (\bibinfo {year} {2011}{\natexlab{b}})}\BibitemShut
  {NoStop}%
\bibitem [{\citenamefont {Ajay}\ and\ \citenamefont
  {Palmer}(1990)}]{palmer1990}%
  \BibitemOpen
  \bibfield  {author} {\bibinfo {author} {\bibnamefont {Ajay}}\ and\ \bibinfo
  {author} {\bibfnamefont {R~G}\ \bibnamefont {Palmer}},\ }\bibfield  {title}
  {\enquote {\bibinfo {title} {Simulation of a toy model with constrained
  dynamics},}\ }\href@noop {} {\bibfield  {journal} {\bibinfo  {journal}
  {Journal of Physics A: Mathematical and General}\ }\textbf {\bibinfo {volume}
  {23}},\ \bibinfo {pages} {2139} (\bibinfo {year} {1990})}\BibitemShut
  {NoStop}%
\bibitem [{\citenamefont {Bohmer}\ \emph {et~al.}(1993)\citenamefont {Bohmer},
  \citenamefont {Ngai}, \citenamefont {Angell},\ and\ \citenamefont
  {Plazek}}]{bohmer1993}%
  \BibitemOpen
  \bibfield  {author} {\bibinfo {author} {\bibfnamefont {R.}~\bibnamefont
  {Bohmer}}, \bibinfo {author} {\bibfnamefont {K.~L.}\ \bibnamefont {Ngai}},
  \bibinfo {author} {\bibfnamefont {C.~A.}\ \bibnamefont {Angell}}, \ and\
  \bibinfo {author} {\bibfnamefont {D.~J.}\ \bibnamefont {Plazek}},\ }\bibfield
   {title} {\enquote {\bibinfo {title} {Nonexponential relaxations in strong
  and fragile glass formers},}\ }\href {\doibase 10.1063/1.466117} {\bibfield
  {journal} {\bibinfo  {journal} {The Journal of Chemical Physics}\ }\textbf
  {\bibinfo {volume} {99}},\ \bibinfo {pages} {4201--4209} (\bibinfo {year}
  {1993})}\BibitemShut {NoStop}%
\bibitem [{\citenamefont {Angell}(1995)}]{angell1995}%
  \BibitemOpen
  \bibfield  {author} {\bibinfo {author} {\bibfnamefont {C~Austen}\
  \bibnamefont {Angell}},\ }\bibfield  {title} {\enquote {\bibinfo {title}
  {Formation of glasses from liquids and biopolymers},}\ }\href@noop {}
  {\bibfield  {journal} {\bibinfo  {journal} {Science}\ }\textbf {\bibinfo
  {volume} {267}},\ \bibinfo {pages} {1924--1935} (\bibinfo {year}
  {1995})}\BibitemShut {NoStop}%
\bibitem [{\citenamefont {Wang}\ \emph {et~al.}(2006)\citenamefont {Wang},
  \citenamefont {Angell},\ and\ \citenamefont {Richert}}]{wang2006}%
  \BibitemOpen
  \bibfield  {author} {\bibinfo {author} {\bibfnamefont {Li-Min}\ \bibnamefont
  {Wang}}, \bibinfo {author} {\bibfnamefont {C.~Austen}\ \bibnamefont
  {Angell}}, \ and\ \bibinfo {author} {\bibfnamefont {Ranko}\ \bibnamefont
  {Richert}},\ }\bibfield  {title} {\enquote {\bibinfo {title} {Fragility and
  thermodynamics in nonpolymeric glass-forming liquids},}\ }\href {\doibase
  10.1063/1.2244551} {\bibfield  {journal} {\bibinfo  {journal} {The Journal of
  Chemical Physics}\ }\textbf {\bibinfo {volume} {125}},\ \bibinfo {pages}
  {074505} (\bibinfo {year} {2006})}\BibitemShut {NoStop}%
\end{thebibliography}%

\end{document}